\documentclass[sigconf]{acmart}
\AtBeginDocument{%
  }

\copyrightyear{2025}
\acmYear{2025}
\setcopyright{none}
\renewcommand\footnotetextcopyrightpermission[1]{}
\acmConference[CHI EA '25]{Extended Abstracts of the CHI Conference on Human Factors in Computing Systems}{April 26-May 1, 2025}{Yokohama, Japan}
\acmBooktitle{Extended Abstracts of the CHI Conference on Human Factors in Computing Systems (CHI EA '25), April 26-May 1, 2025, Yokohama, Japan}\acmDOI{10.1145/3706599.3719802}
\acmISBN{979-8-4007-1395-8/2025/04}

\usepackage{graphicx}
\usepackage{subcaption}
\usepackage{float}
\usepackage{changepage}
\usepackage{pgfplots}
\usepackage{tabularx}
\usepackage{array}
\usepackage{booktabs} 
\usepackage{geometry} 
\usepackage{xcolor}
\usepackage{makecell} 

\begin{document}

\title{Exploring Visual Prompts: Refining Images with Scribbles and Annotations in Generative AI Image Tools}
\author{Hyerim Park}
\affiliation{%
  \institution{BMW Group}
  \city{Munich}
  \country{Germany}}
\affiliation{
  \institution{University of Stuttgart}
  \city{Stuttgart}
  \country{Germany}}
\email{hyerim.park@bmw.de}
\orcid{0009-0006-4877-2255}

\author{Malin Eiband}
\affiliation{%
  \institution{BMW Group}
  \city{Munich}
  \country{Germany}}
\email{malin.eiband@bmw.de}
\orcid{0000-0003-4024-1645}

\author{Andre Luckow}
\affiliation{%
  \institution{BMW Group}
  \city{Munich}
  \country{Germany}}
\affiliation{
  \institution{LMU Munich}
  \city{Munich}
  \country{Germany}}
\email{andre.luckow@bmwgroup.com}
\orcid{0000-0002-1225-4062}

\author{Michael Sedlmair}
\affiliation{%
 \institution{University of Stuttgart}
 \city{Stuttgart}
 \country{Germany}}
\email{michael.sedlmair@visus.uni-stuttgart.de}
\orcid{0000-0001-7048-9292}
\renewcommand{\shortauthors}{Park et al.}

\begin{abstract}
Generative AI (GenAI) tools are increasingly integrated into design workflows. While text prompts remain the primary input method for GenAI image tools, designers often struggle to craft effective ones. Moreover, research has primarily focused on input methods for ideation, with limited attention to refinement tasks. This study explores designers' preferences for three input methods\textemdash text prompts, annotations, and scribbles\textemdash through a preliminary digital paper-based study with seven professional designers. Designers preferred annotations for spatial adjustments and referencing in-image elements, while scribbles were favored for specifying attributes such as shape, size, and position, often combined with other methods. Text prompts excelled at providing detailed descriptions or when designers sought greater GenAI creativity. However, designers expressed concerns about AI misinterpreting annotations and scribbles and the effort needed to create effective text prompts. These insights inform GenAI interface design to better support refinement tasks, align with workflows, and enhance communication with AI systems.
\end{abstract}




\maketitle

\section{Introduction and Background}
Generative AI (GenAI) image generation tools\footnote{In this paper, 'GenAI' refers specifically to generative AI for image generation.} enable users to create high-quality images from simple inputs like text and images, producing outputs comparable to human-created designs. Popular tools include Midjourney \cite{midjourney_midjourney_2024}, DALL-E 3 (integrated into ChatGPT) \cite{dalle_3_dalle_2025}, Adobe Firefly \cite{firefly_adobe_2024}, and DreamStudio (built on Stable Diffusion) \cite{dreamstudio_dreamstudio_2024}. GenAI has been widely adopted across fields such as industrial design \cite{gmeiner_exploring_2023, liu_3dall-e_2023}, graphic design \cite{choi_creativeconnect_2024, ueno_continuous_2021, warner_interactive_2023, jing_layout_2023}, illustration \cite{brade_promptify_2023, liu_opal_2022}, UI/UX design \cite{shi_-stijl_2023, feng_how_2023, lea_user_2023}, and fashion design \cite{wu_styleme_2023}.
With its growing adoption, designing effective interfaces has become a central research focus in Human-Computer Interaction (HCI), particularly in supporting professional designers \cite{choi_creativeconnect_2024, peng_designprompt_2024, chung_promptpaint_2023}. However, qualitative studies highlight that designers often struggle with crafting text prompts, which are currently the dominant input method \cite{gmeiner_exploring_2023, park_we_2024, chen_autospark_2024}. Mismatches between intended designs and generated outputs frequently require corrections, disrupting workflows and diverting attention from core design tasks \cite{takaffoli_generative_2024, zhang_protodreamer_2024}. GenAI's limited contextual understanding and reliance on trial-and-error further complicate iterative workflows, frequently requiring designers to switch tools and increasing cognitive load \cite{tholander_design_2023, park_we_2024}.

To address these challenges, researchers have suggested enhancing text-prompt interfaces with features such as automated keyword suggestions \cite{peng_designprompt_2024, choi_creativeconnect_2024}, revised prompt generation \cite{brade_promptify_2023, son_genquery_2024}, and dropdown menus for interactive refinement \cite{wang_promptcharm_2024}. 
In addition to improving text-based inputs, multimodal approaches incorporating visual inputs have also been explored. Examples include combining text with image uploads, color palettes \cite{peng_designprompt_2024}, or rough sketches \cite{dang_worldsmith_2023}, as well as introducing visual controls like sliders for blending prompt weights or anchors for resizing elements \cite{chung_promptpaint_2023}. However, these visual interactions primarily supplement text prompts rather than serve as primary input methods.

Research on GenAI image tools has mostly focused on ideation and exploration, where these tools support divergent thinking by generating various concepts and inspiring creativity \cite{tholander_design_2023, chen_autospark_2024, choi_creativeconnect_2024, brade_promptify_2023, son_genquery_2024, liang_storydiffusion_2024}. However, refinement tasks, which involve making precise adjustments to better align with user intent, remain underexplored. Our study defines \textit{refinement} as an iterative process of generating and fine-tuning images using GenAI to match a designer's intended vision, which may evolve through interaction with AI-generated outputs. In contrast, \textit{finalization} focuses on meticulous, pixel-perfect modifications aimed at delivering a polished, production-ready result.

Several studies highlight the need for better refinement support in GenAI tools. Users often seek iterative, localized regeneration to modify specific elements \cite{zhang_protodreamer_2024}. As designs evolve, the focus shifts from broad characteristics like style and layout to finer details and spatial composition \cite{dang_worldsmith_2023}. While GenAI tools typically prioritize ideation, refinement tasks demand higher precision and iterative capabilities, which many existing tools lack \cite{liang_storydiffusion_2024}. Current \textit{inpainting} techniques enable users to refine specific image regions by marking areas with brushes, erasers, or stencils, seamlessly blending new content with surrounding elements \cite{yu_generative_2018, dalle_inpainting_editing_2024, midjourney_inpainting_vary_2024}. However, these methods predominantly rely on text prompts to guide edits, which designers often find limiting \cite{dang_worldsmith_2023, park_we_2024, liang_storydiffusion_2024}.

Scribbles and annotations are widely used in design workflows and have potential for refinement tasks. Designers often begin with scribbles, rough sketches, layout drawings, or handwritten annotations to visualize their ideas before working with GenAI \cite{dang_worldsmith_2023}. 
When ideas are vague or underdeveloped, some designers find it challenging to articulate them through text prompts \cite{tholander_design_2023, park_we_2024}, increasing interest in integrating visual inputs with GenAI tools \cite{park_we_2024}.
Some research has differentiated between text-based and visual-based inputs, highlighting their distinct roles in the creative process. For example, one study suggests that sketch-based inputs are more effective for refining and embodying designs in later stages, while text prompts are generally better suited for early ideation \cite{lee_impact_2024}. Similarly, designers often transition from text inputs to image-based inputs for localized variations and then switch back to text prompts to explore new directions \cite{son_genquery_2024}.
To investigate these dynamics, we examine how scribbles and annotations compare to text inputs in refinement tasks using GenAI. Our research is guided by the following question:

\begin{quote}
\textbf{RQ:} How do professional designers perceive and prefer annotations, scribbles, and text-based input methods for different refinement tasks in GenAI image tools?
\end{quote}

We conducted a preliminary user study using a digital paper-based prototype with seven professional designers specializing in UI/UX, automotive interior, graphics, and XR design. Participants completed five types of refinement tasks using text, annotation, and scribble inputs.
Our findings revealed varied preferences among input methods. Annotations were preferred for spatial adjustments (e.g., moving, resizing) and referencing in-image elements (e.g., applying attributes from one object to another). Scribbles were used to specify attributes like shape, position, and size, often complemented by other methods for additional detail. Text inputs were favored for lengthy descriptions involving multiple requirements or tasks that relied on more significant GenAI influence. While experienced users expressed more confidence in text prompts, some participants raised concerns about AI's ability to interpret rough scribbles, handwriting, or custom visual symbols (e.g., arrows, numbering, multi-colored annotations).
These findings highlight the potential of diverse input methods beyond text prompts to support refinement tasks in GenAI image tools, encouraging further exploration to enhance them for later stages of the creative process.

\section{User Study}
\begin{figure*}[!htbp]
    \centering
    \includegraphics[width=1\linewidth,alt={Illustration of the three input methods tested in the study}]{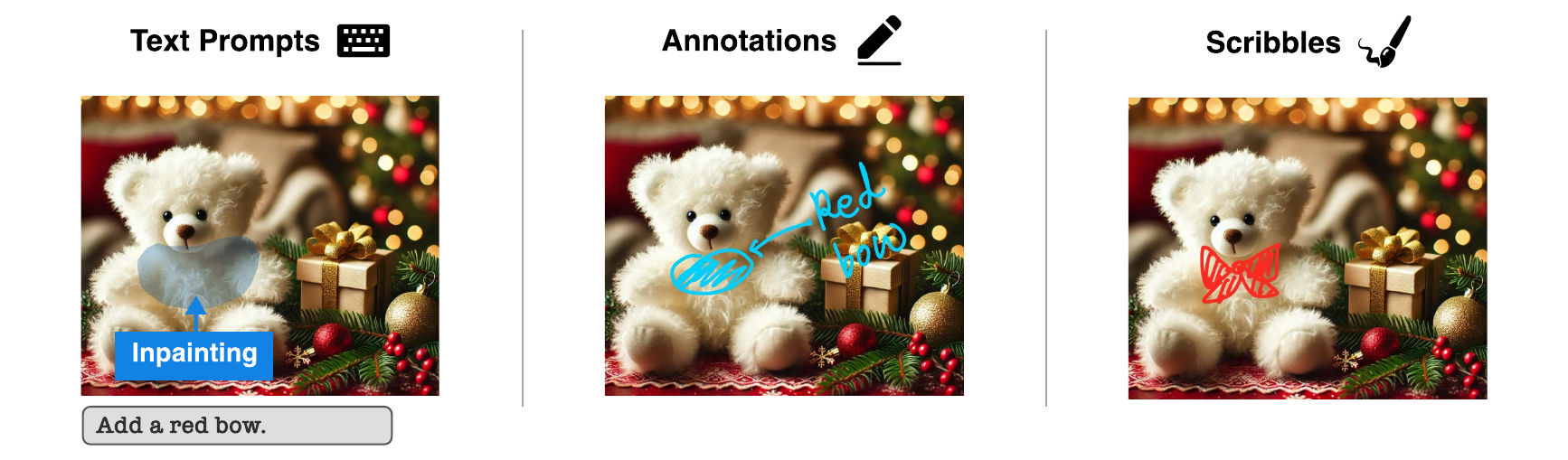}
    \caption{Illustration of the Three Input Methods Tested in the Study: (from left to right) \textbf{text prompts} (typed instructions requiring \textit{inpainting} for area selection), \textbf{annotations} (text or visual symbols on the image, with optional \textit{inpainting} selected by the user), and \textbf{scribbles} (freeform scribbles on the image, with optional \textit{inpainting} selected by the user).}
    \label{fig:three}
    \Description{This figure illustrates the three input methods tested in the study for guiding Generative AI refinements. From left to right, the methods include: (1) Text prompts, where users provide typed instructions that require inpainting for area selection; (2) Annotations, which involve text or visual symbols (such as arrows, circles, and numbering) placed on the image, with optional inpainting selected by the user; and (3) Scribbles, where users create freeform scribbles on the image to indicate modifications, with optional inpainting selected by the user. The figure visually compares these approaches to highlight their differences in user interaction and specificity.}
\end{figure*}

We conducted a preliminary digital paper-based user study to explore designers' preferences for different input methods in refinement tasks using GenAI image tools. Participants interacted with three input methods (see \autoref{fig:three}):
\begin{itemize}
    \item \textbf{Text Prompts:} Typed textual instructions entered via a keyboard.
    \item \textbf{Annotations:} Text or visual symbols (e.g., arrows, circles) added directly to the image using a stylus or mouse.
    \item \textbf{Scribbles:} Freeform sketches drawn directly on the image using a stylus or mouse.
\end{itemize}
\textbf{Marking Edit Areas:} All three methods employed the \textit{inpainting} technique, using a selection brush to mark areas for editing, following standard practices in commercial tools such as Midjourney, Adobe Firefly, and DALL-E \cite{dalle_inpainting_editing_2024, firefly_adobe_2024, midjourney_inpainting_vary_2024}. 
Based on feedback from our pre-tests (N=2), we designed annotations and scribbles to allow users to mark areas directly (e.g., circles, arrows) using the same pen tool, avoiding additional menu navigation. In contrast, text prompts, typed with a keyboard, typically require a selection tool to specify the inpainting area before applying edits. In our user study, most participants used the pen tool for annotations and scribbles rather than the inpainting selection brush.

\subsection{Participants}
Seven designers from an international automotive company participated in the study, recruited via outreach and snowball sampling. Their roles covered UI/UX, graphic, interior, and XR design, with projects extending beyond automotive into broader design contexts. GenAI usage frequency varied: daily (1), 2–3 times weekly (2), weekly (2), monthly (1), and less than monthly (1). Participants primarily used Midjourney, DALL-E 3, and Adobe Firefly. Each in-person session lasted approximately 70 minutes.

\subsection{Procedure}
The study consisted of four phases: (1) introduction and tutorials, (2) pre-tasks, (3) main tasks, and (4) interviews. A think-aloud method was employed, with participants’ verbalizations recorded and supplemented by the experimenter’s observation notes.

\textbf{(1) Introduction and Tutorials:} Participants were briefed on the study's objectives, procedures, and consent process. A tutorial introduced DALL-E 3's text prompting and inpainting capabilities, reflecting the primary methods in most GenAI tools. The three input methods\textemdash text prompts, annotations, and scribbles\textemdash were explained within the context of our simulated settings. Input methods were tested without generating actual outputs to minimize bias from system capabilities (e.g., output quality, latency). Participants selected their preferred input method for each task, with the option to use multiple methods if they favored them equally. For the hardware setup, they could choose between a tablet with a stylus or a laptop with a mouse. During the main tasks, we observed that all participants used a stylus for scribbles and annotations.

\textbf{(2) Pre-Tasks:} Participants completed two pre-tasks to familiarize themselves with the input methods: \textbf{(PT1)} \textit{adding a simple Santa hat to a teddy bear} and \textbf{(PT2)} \textit{modifying the Santa hat into a highly detailed crown.} For each task, they chose their preferred input method(s) and explained how they would guide GenAI to complete it.

\textbf{(3) Main Tasks:} Participants generated several images using DALL-E 3 for presentations, team event posters, or other non-confidential creative projects. After selecting one favorite image, they completed six refinement tasks: \textbf{(T1)} \textit{adding new elements,} \textbf{(T2)} \textit{modifying specific elements (e.g., enhancing complexity),} \textbf{(T3)} \textit{revising materials or surfaces,} \textbf{(T4)} \textit{making structural or spatial adjustments,} \textbf{(T5)} \textit{performing global modifications}, and \textbf{(T6)} \textit{a free task to revise any aspect of their choice.}
The task order was randomized using a Latin square design, with specific task details adapted to each participant's generated images. While the experimenter suggested refinement tasks, participants were encouraged to propose modifications within the task categories to better align with their design goals and enhance their motivation.
    
\textbf{(4) Interviews:} Semi-structured interviews followed the tasks, exploring participants’ experiences with each input method. Discussions covered their advantages, limitations, task-specific preferences, alignment with real-world workflows, and professional challenges. Participants also provided suggestions for improving each input method.
\section{Findings}
An iterative review of participant feedback and observations revealed patterns in input preferences (see \autoref{tab:inputpreferences}). Although the study introduced three distinct input methods, participants frequently combined them in various ways. For instance, scribbles were often used alongside annotations and text prompts to add details and improve clarity during refinement tasks.
While our study primarily focused on refinement, participant feedback also provided insights into the ideation process. We observed that refinement and ideation were not always distinct, as participants frequently shifted between them. Moreover, input methods were not strictly tied to a single stage of the creative process. Participants selected different methods based on the task at hand or the clarity of their design intent. We include these observations to offer a more comprehensive understanding of input preferences.
\begin{table*}[!htbp]
\centering
\footnotesize
\caption{Participants' Input Method Preferences Across Tasks: Multiple methods in a single cell indicate an equal preference for those inputs. Methods in parentheses represent selections for the free task (T6), in which participants chose what to refine.}
\Description{This table shows participants' input method preferences across five different tasks. The table presents methods for each task, including text prompts, annotations, scribbles, and combinations of these methods. In some cells, multiple methods are listed, indicating an equal preference. Task T6 shows the preferred methods for the free task where participants could choose their own input methods.}
\renewcommand{\arraystretch}{1} 
\begin{tabularx}{\textwidth}{@{}p{0.015\textwidth} X X X X X@{}}
\toprule

\textbf{PID} & \textbf{(T1) Add New Elements} & \textbf{(T2) Modify Specific Elements} & \textbf{(T3) Revise Materials or Surfaces} & \textbf{(T4) Structural or Spatial Adjustments} & \textbf{(T5) Global Modifications} \\
\midrule
P1 & \makecell[l]{Text prompts} & \makecell[l]{Scribbles + annotations} & \makecell[l]{Annotations, \\ (T6: Scribbles + annotations)} & \makecell[l]{Scribbles + annotations} & \makecell[l]{Text prompts} \\
\midrule
P2 & \makecell[l]{Scribbles + annotations} & \makecell[l]{Annotations, \\ (T6: Annotations)} & \makecell[l]{Text prompts} & \makecell[l]{Annotations, \\ Scribbles + annotations} & \makecell[l]{Text prompts} \\
\midrule
P3 & \makecell[l]{Scribbles + annotations, \\ Annotations} & \makecell[l]{Annotations, \\ (T6: Annotations)} & \makecell[l]{Text prompts} & \makecell[l]{Scribbles + annotations} & \makecell[l]{Text prompts} \\
\midrule
P4 & \makecell[l]{Scribbles + text prompts} & \makecell[l]{Annotations} & \makecell[l]{Text prompts} & \makecell[l]{Annotations} & \makecell[l]{Text prompts + annotations, \\ (T6: Text prompts)} \\
\midrule
P5 & \makecell[l]{Scribbles + text prompts, \\ Scribbles + text prompts + \\ annotations} & \makecell[l]{Annotations} & \makecell[l]{Text prompts} & \makecell[l]{Annotations, \\ (T6: Annotations)} & \makecell[l]{Text prompts} \\
\midrule
P6 & \makecell[l]{Text prompts} & \makecell[l]{Scribbles + annotations, \\ (T6: Text prompts)} & \makecell[l]{Text prompts} & \makecell[l]{Annotations, \\ Text prompts} & \makecell[l]{Text prompts} \\
\midrule
P7 & \makecell[l]{Text prompts, \\ (T6: Scribbles + text prompts)} & \makecell[l]{Text prompts} & \makecell[l]{Annotations} & \makecell[l]{Annotations} & \makecell[l]{Text prompts} \\
\bottomrule
\end{tabularx}
\label{tab:inputpreferences}
\end{table*}

\subsection{Text Prompts: Effective for Detailed Descriptions and Global Refinements, but Challenging to Craft}

\paragraph{\textbf{Detailed Instructions}}
When designers had a \textbf{clear intent for modification} and a confident \textbf{grasp of how to describe it}, they preferred text inputs for tasks requiring \textbf{lengthy and detailed descriptions} to guide GenAI, especially when specifying \textbf{multiple attributes of a target element}. For example, participants provided prompts such as \textit{"Add a Santa hat in wine-red color, made of velvet, and with sequin decoration"} (P6) or \textit{"Replace her dress with a design that looks like a collaboration with Louis Vuitton designers, incorporating a more luxurious concept"} (P7).

\paragraph{\textbf{Global Modifications}}
Text inputs proved effective for tasks involving \textbf{global changes}, such as modifying an image's \textbf{overall tone or format}. Examples included prompts like \textit{"Remove the yellow tone"} (P6) and \textit{"Change the image ratio to be more square"} (P4). The brush width control in \textit{inpainting} allowed for a more efficient selection of large areas compared to manual annotations (e.g., circling elements) when selecting substantial parts of elements in the image (P5, P6, P7). Designers often used text prompts to specify \textbf{styles}, referencing \textbf{brands, designer names, or design concepts} to achieve desired \textbf{moods or aesthetics} (P6, P7).

\paragraph{\textbf{Creative Flexibility}}
Text inputs were also preferred when participants had \textbf{uncertainty about their design intent or concept} or wanted to leverage \textbf{GenAI's creative input}. Examples included replacing a male figure with a generic female figure for \textit{"a girlfriend look"} (P5), refining kiosk textures to a \textit{"bolder style"} (P3), or adjusting a menu display to a \textit{"more minimal style"} (P1). P3 mentioned relying on AI due to having \textit{"no ideas about kiosk surfaces,"} while P1 appreciated AI's ability to \textit{"find additional details."}

\paragraph{\textbf{Familiarity Among Experienced Users}}
Three designers who regularly used GenAI tools (either daily or 2 to 3 times per week) preferred text prompts more often across tasks in the study. They credited this preference to \textbf{company training} and proactive efforts to \textbf{improve their prompt-crafting skills through online communities and resources}. P7 reflected that their preference might have differed without \textbf{long-term experience}, while P1 remarked, \textit{"I am more familiar with text prompts."}

\paragraph{\textbf{Challenges in Crafting Prompts}}
While effective, \textbf{crafting text prompts remains challenging}. P1 remarked, \textit{"(…) explaining visual ideas in words is not easy."} Others (P2–P6) questioned GenAI's ability to fully interpret their descriptions and noted that creating prompts required \textbf{significant time and effort}. Several participants (P3, P4, P6, P7) anticipated needing \textbf{multiple iterations} to achieve their desired outcomes in real-world scenarios. These challenges align with prior research highlighting the \textbf{cognitive effort} involved in effectively communicating design ideas \cite{takaffoli_generative_2024, park_we_2024}.

\subsection{Annotations: Useful for Spatial and Attribute Referencing, but Concerns About GenAI Interpretation}
\paragraph{\textbf{Spatial Adjustments and In-Image Attribute Referencing}}
Annotations were commonly used for \textbf{spatial adjustments}, such as repositioning a plant or resizing lamps or logos. Designers often used circles and arrows along with brief handwritten notes indicating what GenAI should do (e.g., \textit{"replace"} or \textit{"switch"}) or specifying a target object (e.g., \textit{"ribbon"} or \textit{"dachshund"}). Annotations also effectively supported \textbf{referencing attributes within an image}. Designers used arrows to indicate that a texture from one person's clothing should be applied to another's outfit (P3) or to match the size of two beer glasses by labeling them as \textit{"1"} and \textit{"2"}, then adding a text prompt like \textit{"Make 1 and 2 the same size"} (P5).

\paragraph{\textbf{Directional Guidance}}
For tasks requiring \textbf{directional instructions}, such as lighting direction adjustments or orienting elements (e.g., \textit{"eyes," "audience seats"}), annotations\textemdash particularly symbols like \textbf{dots and arrows}\textemdash were preferred over text prompts for their speed and ease of use. For instance, designers marked lighting directions with arrows, adding instructions like \textit{"Move the sun direction following the arrow"} (P7) or \textit{"Make the eyes look at this point (dot)"} (P5). They noted that they would rely solely on symbols if AI could interpret them correctly (P7). Lighting effect adjustments, such as \textbf{brightening or shadowing surfaces}, were clarified with annotations paired with scribbles, as seen in P1's use of rough strokes combined with annotations like \textit{"brighter"} or \textit{"darker."}

\paragraph{\textbf{Minor Refinements in Later Stages}}
Annotations were preferred for \textbf{small-scale adjustments that required minimal text input} in later refinement stages. For example, P1 and P7 noted that annotations were adequate for minor tasks or a single change, such as \textit{"a bit softer"} or \textit{"more padding"}, due to their simplicity and clarity. In contrast, annotations were less practical in early refinement stages, which often involved more extensive changes, as capturing all details required lengthy descriptions. For instance, P7 considered annotating grass with multiple instructions, such as \textit{"fluffy," "denser with more grass,"} and \textit{"more lighting effects on the surface,"} too labor-intensive by hand, making shorter descriptions prone to vagueness. To add precision, some designers \textbf{combined annotations with scribbles} to specify \textbf{exact dimensions}, such as marking a dog's size, position, and orientation or the spacing between menu buttons.

\paragraph{\textbf{Workflow Integration}}
P6 described \textbf{an envisioned workflow} that involved \textbf{iterative refinements} in the order of text prompts, annotations, and rough scribbles. They typically began with text prompts, like \textit{"Apply a floral pattern to the pants,"} to generate images that allowed for broad AI-driven modifications. Annotations were then used to fine-tune details, marking areas with arrows or circles and adding instructions like \textit{"smaller"} to adjust the pattern size. If further clarification was needed, they conveyed their intent by quickly scribbling the desired size and shape directly onto the image.

\paragraph{\textbf{Visibility and AI Interpretation Challenges}}
Despite their advantages, annotations had limitations. Overlapping annotations on images were seen as disruptive, hindering visual clarity. Without zoom features, \textbf{annotations could quickly obscure the image}, making finer details harder to discern. Some designers (P3-P6) suggested a \textbf{canvas-style layout}, allowing annotations to be written on an empty canvas outside the image. 
Concerns about \textbf{AI's interpretation of handwriting} were also raised (P1, P6). While annotations and scribbles in personal workflows were often \textbf{informal and quick}, participants in the study \textbf{made an effort to write clearly}, knowing their inputs were shared, even though it was not required.

\subsection{Scribbles: Facilitating Simple Edits, but Requiring Complementary Input for Clarity}
\paragraph{\textbf{Precise Goals for Edits Focusing on a Single Attribute}}
Scribbles enabled users to focus on a \textbf{single attribute or a small set of attributes with precision}, especially when \textbf{combined with text prompts or annotations}. They were effective for indicating details such as the \textbf{placement of elements} (e.g., \textit{"a ribbon on the dress"}) (P6), \textbf{shape, size, and angle} (e.g.,\textit{ of "beanies"}) (P5), snow \textbf{distribution} on a mountain (P2), and \textbf{the number and position of players} (P5). Scribbles were also used to resize objects like lamps, dress lengths, and balloons to \textbf{specific dimensions} (P2, P3, P6, P7). This approach was particularly useful when users had \textbf{clear goals and a concrete idea} for modifying specific elements, allowing \textbf{precise marking} to indicate adjustments. For example, P1 employed scribbles to adjust the size and spacing of polka dots on fabric more effectively than using text prompts. Similarly, P4 noted, \textit{"It is easier to draw than explain in text"} and sketched a station design directly on the image, helping expedite the iterative process while providing more control.

\paragraph{\textbf{Universally Recognizable Concepts with Scribbles}}
Scribbles were preferred for \textbf{simple, universally recognized objects} such as \textit{"polka dots," "not very detailed patterns"} (P1), or \textit{"football, bowtie, or simple hat"} (P5). For more complex and unique elements, like a football team logo, users often combined scribbles with additional textual instructions, such as \textit{"Add a Tottenham logo on the pink circle."} As P5 noted, \textit{"It is harder to scribble complex designs as it takes time, and I doubt AI understands them."}

\paragraph{\textbf{Combining with Other Inputs for Clarity}}
While scribbles offered advantages, most users avoided relying on them alone due to concerns about their \textbf{limited detail} and \textbf{uncertainty about AI's interpretation}. Scribbles were often combined with text prompts or annotations to provide context or additional details. For example, P7 said, \textit{"Make the lamp bigger, as I drew it,"} while P4 added, \textit{"Add a flying car station to match my scribble."}
Scribbles also had limitations, particularly for expressing concepts like \textbf{materials and textures} (e.g., \textit{"shiny surfaces"}) (P5), \textbf{abstract ideas} (e.g., \textit{"minimal display"}) (P1), or \textbf{complex instructions that were difficult to convey through scribbles alone} (e.g.,\textit{ "Make the building fit inside the image"}) (P6). Poorly drawn scribbles could mislead the AI, resulting in unintended outcomes. For example, P6 noted that rough sketches, such as a man with an exaggerated head and narrow shoulders, could lead to \textbf{distorted designs}. These concerns highlight the need to \textbf{combine scribbles with text descriptions} to improve clarity and reduce ambiguity.

\paragraph{\textbf{Playful Exploration and Non-Designer Use}}
Some designers (P1, P6, P7) viewed scribbles as \textbf{better suited for non-designers}, enabling experimentation and playful exploration. P7 observed that scribbles work well for \textbf{general users} when precision in \textbf{elements like material, texture, color, or size is not critical.} However, P7 emphasized that professional design work necessitates \textbf{consistent visual quality}.

\paragraph{\textbf{Beneficial for both Ideation and Final Adjustments}}
Some designers found scribbles \textbf{useful for ideation tasks}, especially when instructing \textbf{rough outlines of elements} in an image. Scribbles allowed quick iterations and reduced generation time without requiring precise descriptions of layout and composition. P3 envisioned combining very rough sketches, resembling blocks, with text annotations \textbf{to guide the position, size, and layout of elements in initial images.} Conversely, scribbles were also valued in later stages, with P6 noting that they would be particularly useful for \textbf{fine marking} to make \textbf{precise adjustments} (e.g., subtle refinements in shape). This approach could save time by reducing the need to create text prompts for minor modifications or export work to design tools like Photoshop.
\section{Key Insights and Future Research Directions}
In this section, we present key insights from our findings and outline potential research questions and directions to further explore and enhance GenAI image tool interfaces for design workflows.
Although our study focuses on GenAI in refinement tasks, some directions extend to the broader creative process. As a qualitative preliminary study, our insights also provide useful context for understanding designers' interactions with these tools across workflows.

\subsection{How Can GenAI Dynamically Adapt Input Methods to Meet Diverse User Needs and Contexts?}
Input preferences varied based on the user's need to balance clarity of edits with openness to AI's creative influence. For precise refinements, annotations or scribbles\textemdash often combined with other methods\textemdash were preferred for directly and efficiently specifying changes such as size, position, or shape. 
In contrast, text prompts were favored for open-ended tasks, offering greater room for AI interpretation and creativity, such as global adjustments to styles (e.g., \textit{"minimal,"} \textit{"futuristic"}) or material changes (e.g., \textit{"matte"}).
The balance between control and flexibility shifts across creative phases, projects, and user expertise. Key questions include: How can GenAI adapt to diverse user needs and contexts, providing the proper input methods at the right time? Additionally, how can it enable seamless transitions between methods without disrupting workflows? Exploring these questions could guide the design of more intuitive, adaptable systems.

\subsection{How Can GenAI Interfaces Adapt to Varied Work Setups?}
All participants opted for styluses for annotations and scribbles when provided. Designers who primarily work with styluses were more inclined to envision using these inputs in GenAI, citing compatibility with their workflows. In contrast, those using a mouse or trackpad were less likely to favor these methods due to workflow disruptions, except for one designer who noted that GenAI's iterative nature, requiring multiple edits, made switching tools worthwhile for refinements.
Input preferences may shift in smartphone environments that rely on finger interactions. While not tested, designer interviews and tools like Galaxy AI's sketch-to-image feature \cite{samsung_galaxy_ai_use_2024} suggest finger-based scribbles may appeal to non-designers, enabling creativity with less precision.
Future research could explore how input preferences vary across devices, highlighting user group differences and how GenAI input methods adapt to different setups. For instance, annotations could integrate with a keyboard and mouse through features like sticky notes (e.g., Figma, Miro) or comment tools in document applications (e.g., Word, Google Docs), broadening their applicability across different work setups.

\subsection{How Can GenAI Interfaces Better Interpret User Intent in Annotations?}
Regarding annotation input, participants raised concerns about AI misinterpreting handwriting or simple symbols like arrows, often supplementing them with text prompts for clarity (e.g., \textit{"apply changes to the circle"}). While annotations are typically informal and self-explanatory, participants made extra efforts to ensure AI comprehension during the study.
Key questions arise: How can GenAI better differentiate between annotation types (e.g., symbols, handwriting, labeling) to interpret user intent? What features, such as color coding or numbering, could help users convey instructions more clearly and reduce errors? Integrating conventional features like canvas layouts, zoom, or annotation notes for meta-communication could enhance annotation creation and management. Exploring common annotation symbols and conventions may also guide more intuitive, user-centered GenAI interface design.

\subsection{Should GenAI Interfaces Adapt to Users or Guide Users to Adapt?}
Our study found that designers who frequently used GenAI tools (daily or 2–3 times per week) and engaged in training and online communities showed a stronger preference for text prompts than others. While they acknowledged the challenges of crafting prompts, their skills and confidence improved significantly with experience, easing the initial learning curve.
This observation raises a key question: Should systems accommodate users' existing skills, or should users adapt through training and iterative practice? This balance in current GenAI image tools could be explored in two directions \cite{park_designing_2024}: enhancing support for crafting effective text prompts, encouraging adaptation to text-centered tools, or improving alternative input methods to better support visually oriented designers.

\section{Limitations}
This study has several limitations. First, the small sample size may limit the generalizability of our findings. While we identified patterns, a larger participant pool could reveal additional preferences and nuances. However, as a preliminary study, we believe it provides a valuable starting point for understanding input preferences for refinement. 
Second, the simulated input setting, which isolated input methods from factors like latency or output quality (which could influence trial-and-error cycles), meant that participants did not produce actual outputs. This may have affected their perceptions and limited the depth of their overall experience.
Additionally, this preliminary study observed only a single instance of input creation for specific tasks. We acknowledge that this does not fully reflect the inherently iterative nature of refinement, where designers adjust their inputs after reviewing AI-generated results, influencing their preferred input methods.
Moreover, while we categorized refinement tasks into five main types and included a free task, all free tasks chosen by designers fell within our predefined categories. Additional task types could emerge through long-term real-world observations or larger studies. 
Lastly, while our findings highlight the distinct roles of annotations and scribbles, we also observed some overlap, as designers often used both together for tasks like spatial adjustments and referencing elements. Future research could further explore their unique use cases and how they can be effectively combined in refinement tasks.
Additionally, future work could address the limitations we mentioned by involving a larger sample size to capture broader preferences, conducting system-based tests with generated outputs for a more comprehensive experience (e.g., actual system implementation or Wizard-of-Oz studies), and exploring iterative refinements in real-world contexts to identify overlooked task types.

\section{Conclusion}
This paper explored preferences for annotations, scribbles, and text prompts in GenAI image tools for refinement tasks, addressing gaps in text-centric inputs and the predominant focus of prior research on the ideation phase rather than refinement. Our preliminary digital paper-based study with seven professional designers across five refinement task types revealed that input preferences varied based on the task and user goals.
The findings indicate that annotations were preferred for making spatial adjustments and referencing elements within images. Scribbles, often used alongside other methods, were effective in specifying attributes such as shape, size, and position. Text prompts were favored for tasks requiring detailed instructions or allowing greater creative input from GenAI. However, each method had its drawbacks: annotations and scribbles raised concerns about AI misinterpretation, while crafting effective text prompts demanded significant effort.
Our insights highlight the need for interfaces that balance control and flexibility, considering specific design tasks and diverse work setups while moving beyond traditional text-prompting methods. As AI systems improve in interpreting scribbled or handwritten annotated visual inputs, HCI research could explore effective ways to enhance user intent communication within interfaces. Future research could investigate whether to prioritize the development of text-based prompt creation tools or focus on alternative input methods that better support visually oriented designers and users at different stages of the design process and creative workflow.


\begin{acks}
We thank all participants, as well as the anonymous reviewers, for their time and valuable input. Moreover, we greatly thank GG and Leo for participating in the pre-tests.
For this work, we used ChatGPT-4 Omni to refine phrasing and enhance readability, ensuring all content was self-authored.
\end{acks}

\bibliographystyle{ACM-Reference-Format}
\bibliography{Bib}


\begin{thebibliography}{33}


\ifx \showCODEN    \undefined \def \showCODEN     #1{\unskip}     \fi
\ifx \showDOI      \undefined \def \showDOI       #1{#1}\fi
\ifx \showISBNx    \undefined \def \showISBNx     #1{\unskip}     \fi
\ifx \showISBNxiii \undefined \def \showISBNxiii  #1{\unskip}     \fi
\ifx \showISSN     \undefined \def \showISSN      #1{\unskip}     \fi
\ifx \showLCCN     \undefined \def \showLCCN      #1{\unskip}     \fi
\ifx \shownote     \undefined \def \shownote      #1{#1}          \fi
\ifx \showarticletitle \undefined \def \showarticletitle #1{#1}   \fi
\ifx \showURL      \undefined \def \showURL       {\relax}        \fi
\providecommand\bibfield[2]{#2}
\providecommand\bibinfo[2]{#2}
\providecommand\natexlab[1]{#1}
\providecommand\showeprint[2][]{arXiv:#2}

\bibitem[3(2025)]%
        {dalle_3_dalle_2025}
\bibfield{author}{\bibinfo{person}{DALL·E 3}.} \bibinfo{year}{2025}\natexlab{}.
\newblock \bibinfo{title}{{DALL}·{E} 3}.
\newblock
\newblock
\urldef\tempurl%
\url{https://openai.com/index/dall-e-3/}
\showURL{%
\tempurl}


\bibitem[AI(2024)]%
        {samsung_galaxy_ai_use_2024}
\bibfield{author}{\bibinfo{person}{Samsung~Galaxy AI}.} \bibinfo{year}{2024}\natexlab{}.
\newblock \bibinfo{title}{Use {AI} editing tools in {Gallery} on your {Galaxy} phone or tablet}.
\newblock
\newblock
\urldef\tempurl%
\url{https://www.samsung.com/us/support/answer/ANS10000934/}
\showURL{%
\tempurl}


\bibitem[Brade et~al\mbox{.}(2023)]%
        {brade_promptify_2023}
\bibfield{author}{\bibinfo{person}{Stephen Brade}, \bibinfo{person}{Bryan Wang}, \bibinfo{person}{Mauricio Sousa}, \bibinfo{person}{Sageev Oore}, {and} \bibinfo{person}{Tovi Grossman}.} \bibinfo{year}{2023}\natexlab{}.
\newblock \showarticletitle{Promptify: {Text}-to-{Image} {Generation} through {Interactive} {Prompt} {Exploration} with {Large} {Language} {Models}}. In \bibinfo{booktitle}{\emph{Proceedings of the 36th {Annual} {ACM} {Symposium} on {User} {Interface} {Software} and {Technology}}} \emph{(\bibinfo{series}{{UIST} '23})}. \bibinfo{publisher}{Association for Computing Machinery}, \bibinfo{address}{New York, NY, USA}, \bibinfo{pages}{1--14}.
\newblock
\showISBNx{979-8-4007-0132-0}
\urldef\tempurl%
\url{https://doi.org/10.1145/3586183.3606725}
\showDOI{\tempurl}


\bibitem[Chen et~al\mbox{.}(2024)]%
        {chen_autospark_2024}
\bibfield{author}{\bibinfo{person}{Liuqing Chen}, \bibinfo{person}{Qianzhi Jing}, \bibinfo{person}{Yixin Tsang}, \bibinfo{person}{Qianyi Wang}, \bibinfo{person}{Ruocong Liu}, \bibinfo{person}{Duowei Xia}, \bibinfo{person}{Yunzhan Zhou}, {and} \bibinfo{person}{Lingyun Sun}.} \bibinfo{year}{2024}\natexlab{}.
\newblock \showarticletitle{{AutoSpark}: {Supporting} {Automobile} {Appearance} {Design} {Ideation} with {Kansei} {Engineering} and {Generative} {AI}}. In \bibinfo{booktitle}{\emph{Proceedings of the 37th {Annual} {ACM} {Symposium} on {User} {Interface} {Software} and {Technology}}}. \bibinfo{publisher}{ACM}, \bibinfo{address}{Pittsburgh PA USA}, \bibinfo{pages}{1--19}.
\newblock
\showISBNx{979-8-4007-0628-8}
\urldef\tempurl%
\url{https://doi.org/10.1145/3654777.3676337}
\showDOI{\tempurl}


\bibitem[Choi et~al\mbox{.}(2024)]%
        {choi_creativeconnect_2024}
\bibfield{author}{\bibinfo{person}{DaEun Choi}, \bibinfo{person}{Sumin Hong}, \bibinfo{person}{Jeongeon Park}, \bibinfo{person}{John Joon~Young Chung}, {and} \bibinfo{person}{Juho Kim}.} \bibinfo{year}{2024}\natexlab{}.
\newblock \showarticletitle{{CreativeConnect}: {Supporting} {Reference} {Recombination} for {Graphic} {Design} {Ideation} with {Generative} {AI}}. In \bibinfo{booktitle}{\emph{Proceedings of the {CHI} {Conference} on {Human} {Factors} in {Computing} {Systems}}}. \bibinfo{publisher}{ACM}, \bibinfo{address}{Honolulu HI USA}, \bibinfo{pages}{1--25}.
\newblock
\showISBNx{979-8-4007-0330-0}
\urldef\tempurl%
\url{https://doi.org/10.1145/3613904.3642794}
\showDOI{\tempurl}


\bibitem[Chung and Adar(2023)]%
        {chung_promptpaint_2023}
\bibfield{author}{\bibinfo{person}{John Joon~Young Chung} {and} \bibinfo{person}{Eytan Adar}.} \bibinfo{year}{2023}\natexlab{}.
\newblock \showarticletitle{{PromptPaint}: {Steering} {Text}-to-{Image} {Generation} {Through} {Paint} {Medium}-like {Interactions}}. In \bibinfo{booktitle}{\emph{Proceedings of the 36th {Annual} {ACM} {Symposium} on {User} {Interface} {Software} and {Technology}}}. \bibinfo{publisher}{ACM}, \bibinfo{address}{San Francisco CA USA}, \bibinfo{pages}{1--17}.
\newblock
\showISBNx{979-8-4007-0132-0}
\urldef\tempurl%
\url{https://doi.org/10.1145/3586183.3606777}
\showDOI{\tempurl}


\bibitem[Dang et~al\mbox{.}(2023)]%
        {dang_worldsmith_2023}
\bibfield{author}{\bibinfo{person}{Hai Dang}, \bibinfo{person}{Frederik Brudy}, \bibinfo{person}{George Fitzmaurice}, {and} \bibinfo{person}{Fraser Anderson}.} \bibinfo{year}{2023}\natexlab{}.
\newblock \showarticletitle{{WorldSmith}: {Iterative} and {Expressive} {Prompting} for {World} {Building} with a {Generative} {AI}}. In \bibinfo{booktitle}{\emph{Proceedings of the 36th {Annual} {ACM} {Symposium} on {User} {Interface} {Software} and {Technology}}} \emph{(\bibinfo{series}{{UIST} '23})}. \bibinfo{publisher}{Association for Computing Machinery}, \bibinfo{address}{New York, NY, USA}, \bibinfo{pages}{1--17}.
\newblock
\showISBNx{979-8-4007-0132-0}
\urldef\tempurl%
\url{https://doi.org/10.1145/3586183.3606772}
\showDOI{\tempurl}


\bibitem[Dreamstudio(2024)]%
        {dreamstudio_dreamstudio_2024}
\bibfield{author}{\bibinfo{person}{Dreamstudio}.} \bibinfo{year}{2024}\natexlab{}.
\newblock \bibinfo{title}{{DreamStudio}}.
\newblock
\newblock
\urldef\tempurl%
\url{https://dreamstudio.ai/}
\showURL{%
\tempurl}


\bibitem[Feng et~al\mbox{.}(2023)]%
        {feng_how_2023}
\bibfield{author}{\bibinfo{person}{K.~J.~Kevin Feng}, \bibinfo{person}{Maxwell~James Coppock}, {and} \bibinfo{person}{David~W. McDonald}.} \bibinfo{year}{2023}\natexlab{}.
\newblock \showarticletitle{How {Do} {UX} {Practitioners} {Communicate} {AI} as a {Design} {Material}? {Artifacts}, {Conceptions}, and {Propositions}}. In \bibinfo{booktitle}{\emph{Proceedings of the 2023 {ACM} {Designing} {Interactive} {Systems} {Conference}}}. \bibinfo{publisher}{ACM}, \bibinfo{address}{Pittsburgh PA USA}, \bibinfo{pages}{2263--2280}.
\newblock
\showISBNx{978-1-4503-9893-0}
\urldef\tempurl%
\url{https://doi.org/10.1145/3563657.3596101}
\showDOI{\tempurl}


\bibitem[Firefly(2024)]%
        {firefly_adobe_2024}
\bibfield{author}{\bibinfo{person}{Firefly}.} \bibinfo{year}{2024}\natexlab{}.
\newblock \bibinfo{title}{Adobe {Firefly} - {Free} {Generative} {AI} for creatives}.
\newblock
\newblock
\urldef\tempurl%
\url{https://www.adobe.com/products/firefly.html}
\showURL{%
\tempurl}


\bibitem[Gmeiner et~al\mbox{.}(2023)]%
        {gmeiner_exploring_2023}
\bibfield{author}{\bibinfo{person}{Frederic Gmeiner}, \bibinfo{person}{Humphrey Yang}, \bibinfo{person}{Lining Yao}, \bibinfo{person}{Kenneth Holstein}, {and} \bibinfo{person}{Nikolas Martelaro}.} \bibinfo{year}{2023}\natexlab{}.
\newblock \showarticletitle{Exploring {Challenges} and {Opportunities} to {Support} {Designers} in {Learning} to {Co}-create with {AI}-based {Manufacturing} {Design} {Tools}}. In \bibinfo{booktitle}{\emph{Proceedings of the 2023 {CHI} {Conference} on {Human} {Factors} in {Computing} {Systems}}}. \bibinfo{publisher}{ACM}, \bibinfo{address}{Hamburg Germany}, \bibinfo{pages}{1--20}.
\newblock
\showISBNx{978-1-4503-9421-5}
\urldef\tempurl%
\url{https://doi.org/10.1145/3544548.3580999}
\showDOI{\tempurl}


\bibitem[Inpainting(2024a)]%
        {dalle_inpainting_editing_2024}
\bibfield{author}{\bibinfo{person}{DALL·E Inpainting}.} \bibinfo{year}{2024}\natexlab{a}.
\newblock \bibinfo{title}{Editing your images with {DALL}·{E} {\textbar} {OpenAI} {Help} {Center}}.
\newblock
\newblock
\urldef\tempurl%
\url{https://help.openai.com/en/articles/9055440-editing-your-images-with-dall-e}
\showURL{%
\tempurl}


\bibitem[Inpainting(2024b)]%
        {midjourney_inpainting_vary_2024}
\bibfield{author}{\bibinfo{person}{Midjourney Inpainting}.} \bibinfo{year}{2024}\natexlab{b}.
\newblock \bibinfo{title}{Vary {Region} + {Remix}}.
\newblock
\newblock
\urldef\tempurl%
\url{https://docs.midjourney.com/docs/vary-region}
\showURL{%
\tempurl}


\bibitem[Jing et~al\mbox{.}(2023)]%
        {jing_layout_2023}
\bibfield{author}{\bibinfo{person}{Qianzhi Jing}, \bibinfo{person}{Tingting Zhou}, \bibinfo{person}{Yixin Tsang}, \bibinfo{person}{Liuqing Chen}, \bibinfo{person}{Lingyun Sun}, \bibinfo{person}{Yankun Zhen}, {and} \bibinfo{person}{Yichun Du}.} \bibinfo{year}{2023}\natexlab{}.
\newblock \showarticletitle{Layout {Generation} for {Various} {Scenarios} in {Mobile} {Shopping} {Applications}}. In \bibinfo{booktitle}{\emph{Proceedings of the 2023 {CHI} {Conference} on {Human} {Factors} in {Computing} {Systems}}} \emph{(\bibinfo{series}{{CHI} '23})}. \bibinfo{publisher}{Association for Computing Machinery}, \bibinfo{address}{New York, NY, USA}, \bibinfo{pages}{1--18}.
\newblock
\showISBNx{978-1-4503-9421-5}
\urldef\tempurl%
\url{https://doi.org/10.1145/3544548.3581446}
\showDOI{\tempurl}


\bibitem[Lea et~al\mbox{.}(2023)]%
        {lea_user_2023}
\bibfield{author}{\bibinfo{person}{Colin Lea}, \bibinfo{person}{Zifang Huang}, \bibinfo{person}{Jaya Narain}, \bibinfo{person}{Lauren Tooley}, \bibinfo{person}{Dianna Yee}, \bibinfo{person}{Dung~Tien Tran}, \bibinfo{person}{Panayiotis Georgiou}, \bibinfo{person}{Jeffrey~P Bigham}, {and} \bibinfo{person}{Leah Findlater}.} \bibinfo{year}{2023}\natexlab{}.
\newblock \showarticletitle{From {User} {Perceptions} to {Technical} {Improvement}: {Enabling} {People} {Who} {Stutter} to {Better} {Use} {Speech} {Recognition}}. In \bibinfo{booktitle}{\emph{Proceedings of the 2023 {CHI} {Conference} on {Human} {Factors} in {Computing} {Systems}}}. \bibinfo{publisher}{ACM}, \bibinfo{address}{Hamburg Germany}, \bibinfo{pages}{1--16}.
\newblock
\showISBNx{978-1-4503-9421-5}
\urldef\tempurl%
\url{https://doi.org/10.1145/3544548.3581224}
\showDOI{\tempurl}


\bibitem[Lee et~al\mbox{.}(2024)]%
        {lee_impact_2024}
\bibfield{author}{\bibinfo{person}{Seung~Won Lee}, \bibinfo{person}{Tae~Hee Jo}, \bibinfo{person}{Semin Jin}, \bibinfo{person}{Jiin Choi}, \bibinfo{person}{Kyungwon Yun}, \bibinfo{person}{Sergio Bromberg}, \bibinfo{person}{Seonghoon Ban}, {and} \bibinfo{person}{Kyung~Hoon Hyun}.} \bibinfo{year}{2024}\natexlab{}.
\newblock \showarticletitle{The {Impact} of {Sketch}-guided vs. {Prompt}-guided {3D} {Generative} {AIs} on the {Design} {Exploration} {Process}}. In \bibinfo{booktitle}{\emph{Proceedings of the {CHI} {Conference} on {Human} {Factors} in {Computing} {Systems}}}. \bibinfo{publisher}{ACM}, \bibinfo{address}{Honolulu HI USA}, \bibinfo{pages}{1--18}.
\newblock
\showISBNx{979-8-4007-0330-0}
\urldef\tempurl%
\url{https://doi.org/10.1145/3613904.3642218}
\showDOI{\tempurl}


\bibitem[Liang et~al\mbox{.}(2024)]%
        {liang_storydiffusion_2024}
\bibfield{author}{\bibinfo{person}{Zhaohui Liang}, \bibinfo{person}{Xiaoyu Zhang}, \bibinfo{person}{Kevin Ma}, \bibinfo{person}{Zhao Liu}, \bibinfo{person}{Xipei Ren}, \bibinfo{person}{Kosa Goucher-Lambert}, {and} \bibinfo{person}{Can Liu}.} \bibinfo{year}{2024}\natexlab{}.
\newblock \bibinfo{title}{{StoryDiffusion}: {How} to {Support} {UX} {Storyboarding} {With} {Generative}-{AI}}.
\newblock
\newblock
\urldef\tempurl%
\url{https://doi.org/10.48550/arXiv.2407.07672}
\showDOI{\tempurl}
\newblock
\shownote{arXiv:2407.07672}.


\bibitem[Liu et~al\mbox{.}(2022)]%
        {liu_opal_2022}
\bibfield{author}{\bibinfo{person}{Vivian Liu}, \bibinfo{person}{Han Qiao}, {and} \bibinfo{person}{Lydia Chilton}.} \bibinfo{year}{2022}\natexlab{}.
\newblock \showarticletitle{Opal: {Multimodal} {Image} {Generation} for {News} {Illustration}}. In \bibinfo{booktitle}{\emph{Proceedings of the 35th {Annual} {ACM} {Symposium} on {User} {Interface} {Software} and {Technology}}} \emph{(\bibinfo{series}{{UIST} '22})}. \bibinfo{publisher}{Association for Computing Machinery}, \bibinfo{address}{New York, NY, USA}, \bibinfo{pages}{1--17}.
\newblock
\showISBNx{978-1-4503-9320-1}
\urldef\tempurl%
\url{https://doi.org/10.1145/3526113.3545621}
\showDOI{\tempurl}


\bibitem[Liu et~al\mbox{.}(2023)]%
        {liu_3dall-e_2023}
\bibfield{author}{\bibinfo{person}{Vivian Liu}, \bibinfo{person}{Jo Vermeulen}, \bibinfo{person}{George Fitzmaurice}, {and} \bibinfo{person}{Justin Matejka}.} \bibinfo{year}{2023}\natexlab{}.
\newblock \showarticletitle{{3DALL}-{E}: {Integrating} {Text}-to-{Image} {AI} in {3D} {Design} {Workflows}}. In \bibinfo{booktitle}{\emph{Proceedings of the 2023 {ACM} {Designing} {Interactive} {Systems} {Conference}}}. \bibinfo{publisher}{ACM}, \bibinfo{address}{Pittsburgh PA USA}, \bibinfo{pages}{1955--1977}.
\newblock
\showISBNx{978-1-4503-9893-0}
\urldef\tempurl%
\url{https://doi.org/10.1145/3563657.3596098}
\showDOI{\tempurl}


\bibitem[Midjourney(2024)]%
        {midjourney_midjourney_2024}
\bibfield{author}{\bibinfo{person}{Midjourney}.} \bibinfo{year}{2024}\natexlab{}.
\newblock \bibinfo{title}{Midjourney}.
\newblock
\newblock
\urldef\tempurl%
\url{https://www.midjourney.com/home?callbackUrl=%2Fexplore}
\showURL{%
\tempurl}


\bibitem[Park and Eiband(2024)]%
        {park_designing_2024}
\bibfield{author}{\bibinfo{person}{Hyerim Park} {and} \bibinfo{person}{Malin Eiband}.} \bibinfo{year}{2024}\natexlab{}.
\newblock \showarticletitle{Designing for {Visual} {Thinkers}: {Overcoming} {Text}-{Centric} {Limitations} in {GenAI} {Tools}}.
\newblock
\urldef\tempurl%
\url{https://doi.org/10.5281/zenodo.14186390}
\showDOI{\tempurl}


\bibitem[Park et~al\mbox{.}(2024)]%
        {park_we_2024}
\bibfield{author}{\bibinfo{person}{Hyerim Park}, \bibinfo{person}{Joscha Eirich}, \bibinfo{person}{Andre Luckow}, {and} \bibinfo{person}{Michael Sedlmair}.} \bibinfo{year}{2024}\natexlab{}.
\newblock \showarticletitle{"{We} {Are} {Visual} {Thinkers}, {Not} {Verbal} {Thinkers}!": {A} {Thematic} {Analysis} of {How} {Professional} {Designers} {Use} {Generative} {AI} {Image} {Generation} {Tools}}. In \bibinfo{booktitle}{\emph{Nordic {Conference} on {Human}-{Computer} {Interaction}}}. \bibinfo{publisher}{ACM}, \bibinfo{address}{Uppsala Sweden}, \bibinfo{pages}{1--14}.
\newblock
\showISBNx{979-8-4007-0966-1}
\urldef\tempurl%
\url{https://doi.org/10.1145/3679318.3685370}
\showDOI{\tempurl}


\bibitem[Peng et~al\mbox{.}(2024)]%
        {peng_designprompt_2024}
\bibfield{author}{\bibinfo{person}{Xiaohan Peng}, \bibinfo{person}{Janin Koch}, {and} \bibinfo{person}{Wendy~E. Mackay}.} \bibinfo{year}{2024}\natexlab{}.
\newblock \showarticletitle{{DesignPrompt}: {Using} {Multimodal} {Interaction} for {Design} {Exploration} with {Generative} {AI}}. In \bibinfo{booktitle}{\emph{Designing {Interactive} {Systems} {Conference}}}. \bibinfo{publisher}{ACM}, \bibinfo{address}{IT University of Copenhagen Denmark}, \bibinfo{pages}{804--818}.
\newblock
\showISBNx{979-8-4007-0583-0}
\urldef\tempurl%
\url{https://doi.org/10.1145/3643834.3661588}
\showDOI{\tempurl}


\bibitem[Shi et~al\mbox{.}(2023)]%
        {shi_-stijl_2023}
\bibfield{author}{\bibinfo{person}{Xinyu Shi}, \bibinfo{person}{Ziqi Zhou}, \bibinfo{person}{Jing~Wen Zhang}, \bibinfo{person}{Ali Neshati}, \bibinfo{person}{Anjul~Kumar Tyagi}, \bibinfo{person}{Ryan Rossi}, \bibinfo{person}{Shunan Guo}, \bibinfo{person}{Fan Du}, {and} \bibinfo{person}{Jian Zhao}.} \bibinfo{year}{2023}\natexlab{}.
\newblock \showarticletitle{De-{Stijl}: {Facilitating} {Graphics} {Design} with {Interactive} {2D} {Color} {Palette} {Recommendation}}. In \bibinfo{booktitle}{\emph{Proceedings of the 2023 {CHI} {Conference} on {Human} {Factors} in {Computing} {Systems}}} \emph{(\bibinfo{series}{{CHI} '23})}. \bibinfo{publisher}{Association for Computing Machinery}, \bibinfo{address}{New York, NY, USA}, \bibinfo{pages}{1--19}.
\newblock
\showISBNx{978-1-4503-9421-5}
\urldef\tempurl%
\url{https://doi.org/10.1145/3544548.3581070}
\showDOI{\tempurl}


\bibitem[Son et~al\mbox{.}(2024)]%
        {son_genquery_2024}
\bibfield{author}{\bibinfo{person}{Kihoon Son}, \bibinfo{person}{DaEun Choi}, \bibinfo{person}{Tae~Soo Kim}, \bibinfo{person}{Young-Ho Kim}, {and} \bibinfo{person}{Juho Kim}.} \bibinfo{year}{2024}\natexlab{}.
\newblock \showarticletitle{{GenQuery}: {Supporting} {Expressive} {Visual} {Search} with {Generative} {Models}}. In \bibinfo{booktitle}{\emph{Proceedings of the {CHI} {Conference} on {Human} {Factors} in {Computing} {Systems}}} \emph{(\bibinfo{series}{{CHI} '24})}. \bibinfo{publisher}{Association for Computing Machinery}, \bibinfo{address}{New York, NY, USA}, \bibinfo{pages}{1--19}.
\newblock
\showISBNx{979-8-4007-0330-0}
\urldef\tempurl%
\url{https://doi.org/10.1145/3613904.3642847}
\showDOI{\tempurl}


\bibitem[Takaffoli et~al\mbox{.}(2024)]%
        {takaffoli_generative_2024}
\bibfield{author}{\bibinfo{person}{Macy Takaffoli}, \bibinfo{person}{Sijia Li}, {and} \bibinfo{person}{Ville Mäkelä}.} \bibinfo{year}{2024}\natexlab{}.
\newblock \showarticletitle{Generative {AI} in {User} {Experience} {Design} and {Research}: {How} {Do} {UX} {Practitioners}, {Teams}, and {Companies} {Use} {GenAI} in {Industry}?}. In \bibinfo{booktitle}{\emph{Designing {Interactive} {Systems} {Conference}}}. \bibinfo{publisher}{ACM}, \bibinfo{address}{IT University of Copenhagen Denmark}, \bibinfo{pages}{1579--1593}.
\newblock
\showISBNx{979-8-4007-0583-0}
\urldef\tempurl%
\url{https://doi.org/10.1145/3643834.3660720}
\showDOI{\tempurl}


\bibitem[Tholander and Jonsson(2023)]%
        {tholander_design_2023}
\bibfield{author}{\bibinfo{person}{Jakob Tholander} {and} \bibinfo{person}{Martin Jonsson}.} \bibinfo{year}{2023}\natexlab{}.
\newblock \showarticletitle{Design {Ideation} with {AI} - {Sketching}, {Thinking} and {Talking} with {Generative} {Machine} {Learning} {Models}}. In \bibinfo{booktitle}{\emph{Proceedings of the 2023 {ACM} {Designing} {Interactive} {Systems} {Conference}}} \emph{(\bibinfo{series}{{DIS} '23})}. \bibinfo{publisher}{Association for Computing Machinery}, \bibinfo{address}{New York, NY, USA}, \bibinfo{pages}{1930--1940}.
\newblock
\showISBNx{978-1-4503-9893-0}
\urldef\tempurl%
\url{https://doi.org/10.1145/3563657.3596014}
\showDOI{\tempurl}


\bibitem[Ueno and Satoh(2021)]%
        {ueno_continuous_2021}
\bibfield{author}{\bibinfo{person}{Michihiko Ueno} {and} \bibinfo{person}{Shin’ichi Satoh}.} \bibinfo{year}{2021}\natexlab{}.
\newblock \showarticletitle{Continuous and {Gradual} {Style} {Changes} of {Graphic} {Designs} with {Generative} {Model}}. In \bibinfo{booktitle}{\emph{Proceedings of the 26th {International} {Conference} on {Intelligent} {User} {Interfaces}}} \emph{(\bibinfo{series}{{IUI} '21})}. \bibinfo{publisher}{Association for Computing Machinery}, \bibinfo{address}{New York, NY, USA}, \bibinfo{pages}{280--289}.
\newblock
\showISBNx{978-1-4503-8017-1}
\urldef\tempurl%
\url{https://doi.org/10.1145/3397481.3450666}
\showDOI{\tempurl}


\bibitem[Wang et~al\mbox{.}(2024)]%
        {wang_promptcharm_2024}
\bibfield{author}{\bibinfo{person}{Zhijie Wang}, \bibinfo{person}{Yuheng Huang}, \bibinfo{person}{Da Song}, \bibinfo{person}{Lei Ma}, {and} \bibinfo{person}{Tianyi Zhang}.} \bibinfo{year}{2024}\natexlab{}.
\newblock \showarticletitle{{PromptCharm}: {Text}-to-{Image} {Generation} through {Multi}-modal {Prompting} and {Refinement}}. In \bibinfo{booktitle}{\emph{Proceedings of the {CHI} {Conference} on {Human} {Factors} in {Computing} {Systems}}}. \bibinfo{publisher}{ACM}, \bibinfo{address}{Honolulu HI USA}, \bibinfo{pages}{1--21}.
\newblock
\showISBNx{979-8-4007-0330-0}
\urldef\tempurl%
\url{https://doi.org/10.1145/3613904.3642803}
\showDOI{\tempurl}


\bibitem[Warner et~al\mbox{.}(2023)]%
        {warner_interactive_2023}
\bibfield{author}{\bibinfo{person}{Jeremy Warner}, \bibinfo{person}{Kyu~Won Kim}, {and} \bibinfo{person}{Bjoern Hartmann}.} \bibinfo{year}{2023}\natexlab{}.
\newblock \showarticletitle{Interactive {Flexible} {Style} {Transfer} for {Vector} {Graphics}}. In \bibinfo{booktitle}{\emph{Proceedings of the 36th {Annual} {ACM} {Symposium} on {User} {Interface} {Software} and {Technology}}} \emph{(\bibinfo{series}{{UIST} '23})}. \bibinfo{publisher}{Association for Computing Machinery}, \bibinfo{address}{New York, NY, USA}, \bibinfo{pages}{1--14}.
\newblock
\showISBNx{979-8-4007-0132-0}
\urldef\tempurl%
\url{https://doi.org/10.1145/3586183.3606751}
\showDOI{\tempurl}


\bibitem[Wu et~al\mbox{.}(2023)]%
        {wu_styleme_2023}
\bibfield{author}{\bibinfo{person}{Di Wu}, \bibinfo{person}{Zhiwang Yu}, \bibinfo{person}{Nan Ma}, \bibinfo{person}{Jianan Jiang}, \bibinfo{person}{Yuetian Wang}, \bibinfo{person}{Guixiang Zhou}, \bibinfo{person}{Hanhui Deng}, {and} \bibinfo{person}{Yi Li}.} \bibinfo{year}{2023}\natexlab{}.
\newblock \showarticletitle{{StyleMe}: {Towards} {Intelligent} {Fashion} {Generation} with {Designer} {Style}}. In \bibinfo{booktitle}{\emph{Proceedings of the 2023 {CHI} {Conference} on {Human} {Factors} in {Computing} {Systems}}} \emph{(\bibinfo{series}{{CHI} '23})}. \bibinfo{publisher}{Association for Computing Machinery}, \bibinfo{address}{New York, NY, USA}, \bibinfo{pages}{1--16}.
\newblock
\showISBNx{978-1-4503-9421-5}
\urldef\tempurl%
\url{https://doi.org/10.1145/3544548.3581377}
\showDOI{\tempurl}


\bibitem[Yu et~al\mbox{.}(2018)]%
        {yu_generative_2018}
\bibfield{author}{\bibinfo{person}{Jiahui Yu}, \bibinfo{person}{Zhe Lin}, \bibinfo{person}{Jimei Yang}, \bibinfo{person}{Xiaohui Shen}, \bibinfo{person}{Xin Lu}, {and} \bibinfo{person}{Thomas~S. Huang}.} \bibinfo{year}{2018}\natexlab{}.
\newblock \bibinfo{title}{Generative {Image} {Inpainting} with {Contextual} {Attention}}.
\newblock
\newblock
\urldef\tempurl%
\url{https://doi.org/10.48550/arXiv.1801.07892}
\showDOI{\tempurl}
\newblock
\shownote{arXiv:1801.07892 [cs]}.


\bibitem[Zhang et~al\mbox{.}(2024)]%
        {zhang_protodreamer_2024}
\bibfield{author}{\bibinfo{person}{Hongbo Zhang}, \bibinfo{person}{Pei Chen}, \bibinfo{person}{Xuelong Xie}, \bibinfo{person}{Chaoyi Lin}, \bibinfo{person}{Lianyan Liu}, \bibinfo{person}{Zhuoshu Li}, \bibinfo{person}{Weitao You}, {and} \bibinfo{person}{Lingyun Sun}.} \bibinfo{year}{2024}\natexlab{}.
\newblock \showarticletitle{{ProtoDreamer}: {A} {Mixed}-prototype {Tool} {Combining} {Physical} {Model} and {Generative} {AI} to {Support} {Conceptual} {Design}}. In \bibinfo{booktitle}{\emph{Proceedings of the 37th {Annual} {ACM} {Symposium} on {User} {Interface} {Software} and {Technology}}} \emph{(\bibinfo{series}{{UIST} '24})}. \bibinfo{publisher}{Association for Computing Machinery}, \bibinfo{address}{New York, NY, USA}, \bibinfo{pages}{1--18}.
\newblock
\showISBNx{979-8-4007-0628-8}
\urldef\tempurl%
\url{https://doi.org/10.1145/3654777.3676399}
\showDOI{\tempurl}


\end{thebibliography}

\appendix
\onecolumn
\section{User Input Examples: Guiding GenAI for Refinements}
\label{app:examples}
This appendix provides examples from the user study, illustrating how participants used their preferred input methods\textemdash text prompts, annotations, and scribbles\textemdash to guide GenAI image tools in various refinement tasks. These methods were not strictly separate; at times, participants considered combining multiple inputs within a single task, thinking that this approach might lead to more effective outcomes.

\subsection{Text Prompts}
\begin{figure}[!htbp]
\includegraphics[width=1\linewidth, alt={Examples of text prompt inputs demonstrating different uses of inpainting and direct text instructions for image refinement.}]{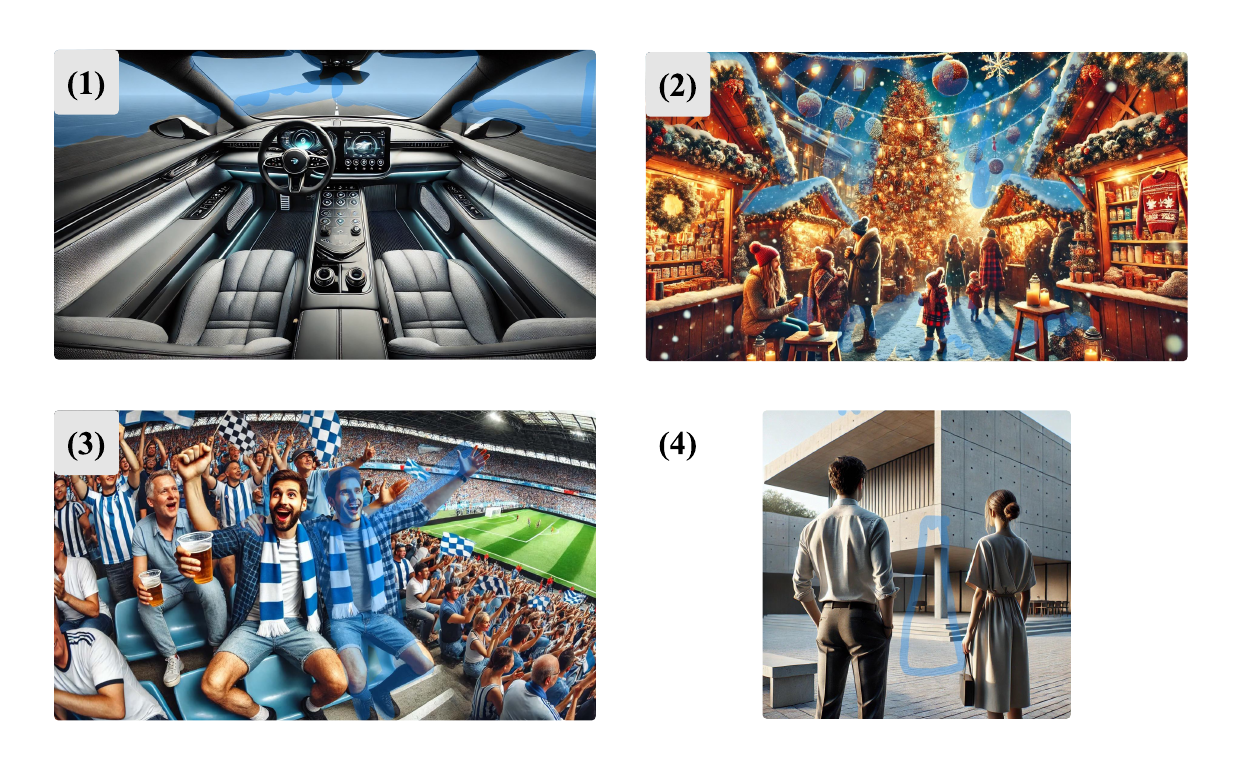}
\caption{\textbf{Text Prompt Input Examples:} (1) and (2) involve \textbf{inpainting} (highlighted in blue) applied to large areas for \textbf{background changes}, such as adding a forest (\textit{"Add a forest as a background outside the car"}) or increasing snowfall (\textit{"Make the snowfall bigger and heavier"}). 
(3) A participant used inpainting to modify a man into a girlfriend figure with the text prompt, \textit{"Change this man into a woman with a girlfriend look."} The girlfriend figure allowed room for AI's creative interpretation, as the prompt required less precise instructions. 
(4) A participant marked an area and provided the text prompt, \textit{"Add a Christmas tree decorated with ornaments and lighting. The decoration should be modern and chic, with yellow and white lights. It should match the building behind."} The instructions were relatively lengthy and detailed.
Some participants relied solely on text prompts without inpainting to adjust the \textbf{overall mood} of the image, such as \textbf{removing the yellow tone entirely}.}
\label{fig:text_examples}
\Description{This figure presents examples of text prompt inputs showcasing various uses of inpainting and direct text instructions for image refinement. The figure contains four examples: (1) and (2) demonstrate inpainting applied to large areas to modify the background, such as adding a forest or intensifying snowfall. (3) illustrates a participant using inpainting to transform a man into a girlfriend figure, allowing AI creative freedom. (4) shows a detailed prompt instructing the addition of a modern Christmas tree with specific decorations. Some participants adjusted the overall mood of images solely through text prompts, such as removing a yellow tone.}
\end{figure}
\clearpage

\subsection{Annotations}
\begin{figure}[!htbp]
    \includegraphics[width=1\linewidth, alt={Examples of annotation-based inputs using text, arrows, circles, numbering, and other visual symbols to guide GenAI modifications in images.}]{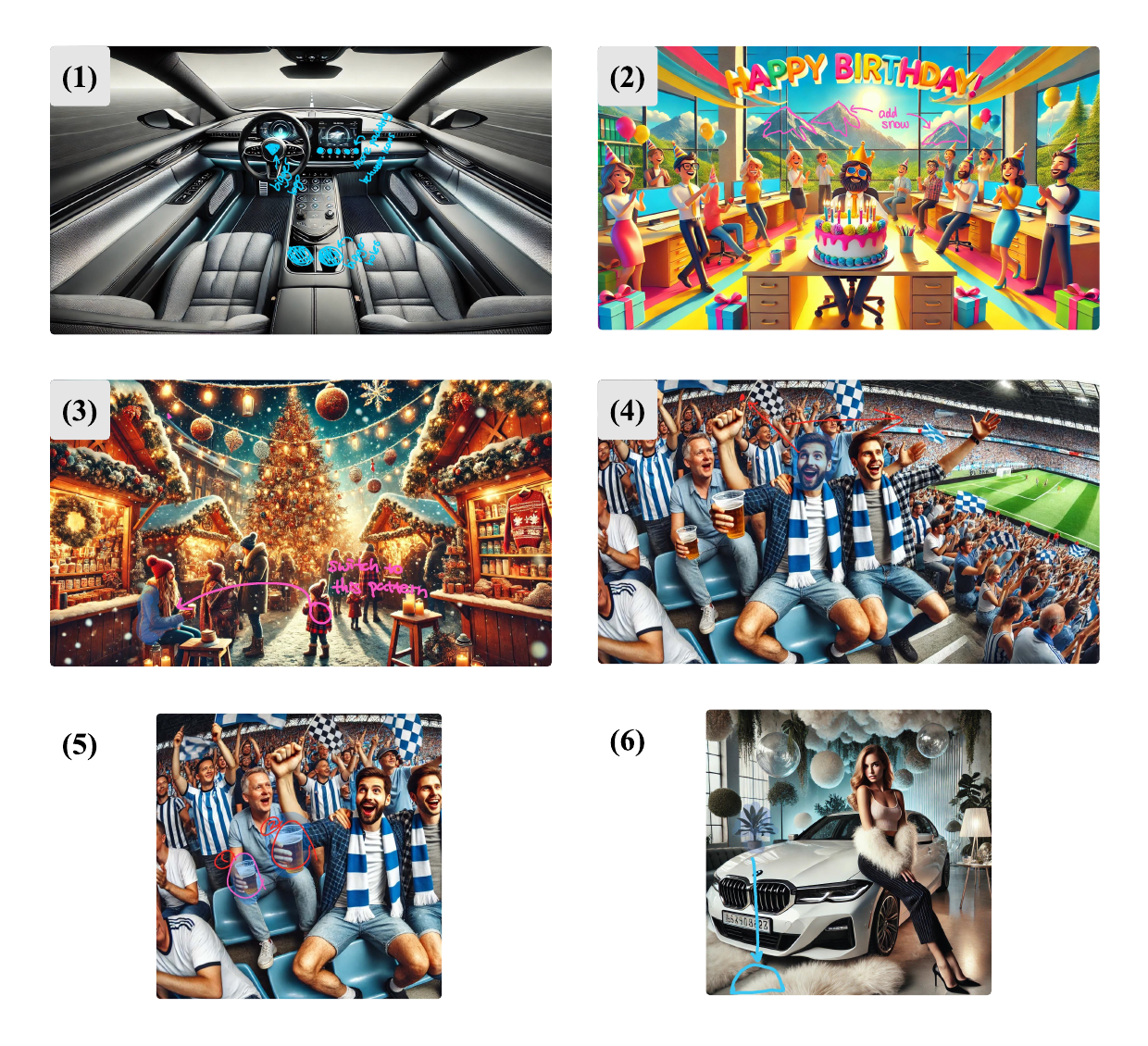}
\caption{\textbf{Annotation Input Examples:} Annotations involve text or visual symbols (e.g., circles, arrows, lines, numbering), with arrows and circles being the most commonly used. \textbf{At times, annotations were combined with scribbles and text prompts.} 
(1) A participant used scribbles to indicate the size of \textbf{a logo on the steering wheel or buttons on the display}, paired with text annotations like \textit{"bigger"} or \textit{"more padding."} 
(2) Scribbles indicated snow distribution on mountains, clarified by a text annotation reading \textit{"Add snow."} 
(3) An arrow and text annotation (\textit{"Switch to this pattern"}) were used to apply a \textbf{checkered pattern from one person’s clothing to another’s} in the image. 
(4) A participant used an arrow and dot to indicate \textbf{a change in eye direction}, with a text prompt stating \textit{"Change the eyes to look in the direction of the dots."}
(5) \textbf{Numbering} (e.g., \textit{"1"} and \textit{"2"}) labeled two beer glasses, with a text prompt stating, \textit{"Make 1 and 2 the same size."} 
(6) Visual symbols, including \textbf{an arrow and a circle}, showed how to \textbf{move a plant to a new position }in the image.}
  \label{fig:annotations}
\Description{This figure presents examples of annotation-based inputs that use text, arrows, circles, numbering, and other visual symbols to guide Generative AI modifications in images. The annotations include: (1) scribbles indicating the desired size adjustments for a logo on a steering wheel or buttons on a display, accompanied by text annotations like "bigger" or "more padding"; (2) scribbles marking snow distribution on mountains, clarified with the text "Add snow"; (3) an arrow and text annotation ("Switch to this pattern") directing the transfer of a checkered pattern from one person’s clothing to another; (4) an arrow and dot marking a change in eye direction, with a text prompt instructing "Change the eyes to look in the direction of the dots"; (5) numbering (e.g., "1" and "2") labeling two beer glasses, with a prompt stating "Make 1 and 2 the same size"; and (6) an arrow and a circle indicating how to reposition a plant within the image.}
\end{figure}
\clearpage

\subsection{Scribbles}
\begin{figure}[!htbp]
    \includegraphics[width=1\linewidth, alt={Examples of scribble-based inputs used to guide GenAI refinements, demonstrating how users indicated shape, size, lighting, positioning, and object details.}]{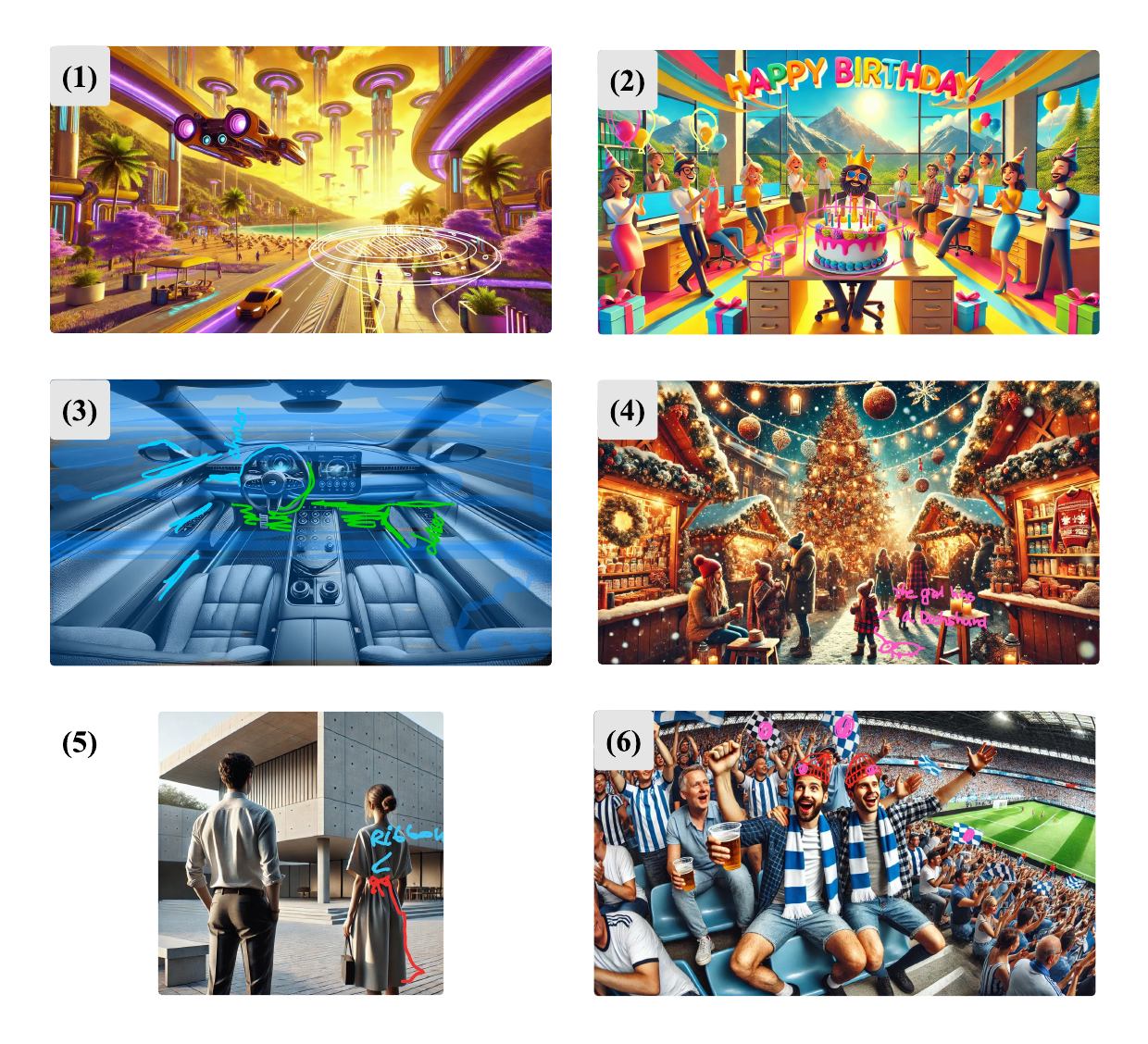}
\caption{\textbf{Scribble Input Examples:} Most users combined scribbles with text prompts or annotations to add details. 
(1) A scribble depicted the \textbf{shape, size, and rough design} of a flying car station to be added. 
(2) A participant used scribbles to specify the intended \textbf{size of a cake, a balloon, and a new element (a cup) on a desk} without additional input. 
(3) Scribbles highlighted \textbf{specific regions of a car for lighting effect adjustments}, paired with annotations specifying areas to brighten (\textit{in blue}) and darken (\textit{in green}). 
(4) A participant used rough scribbles to indicate the \textbf{size and orientation of a dachshund}, complemented by an annotation labeling it as "\textit{dachshund}." 
(5) Scribbles showed the \textbf{desired position of a ribbon and modifications to a skirt shape}, with an annotation specifying "\textit{ribbon}."
(6) A scribble outlined a \textbf{beanie design without a pompom}, while annotations and text prompts detailed adding a “\textit{Tottenham logo}” to the pink-circled area.}
  \label{fig:scribbles}
\Description{This figure presents examples of scribble-based inputs used to guide Generative AI refinements, demonstrating how users indicated shape, size, lighting, positioning, and object details. The examples include: (1) a scribble outlining the shape, size, and rough design of a flying car station to be added; (2) scribbles specifying the intended size of a cake, a balloon, and a new element (a cup) on a desk without additional input; (3) scribbles highlighting specific regions of a car for lighting effect adjustments, with annotations marking areas to brighten (in blue) and darken (in green); (4) rough scribbles indicating the size and orientation of a dachshund, complemented by an annotation labeling it as "dachshund"; (5) scribbles showing the desired position of a ribbon and modifications to a skirt shape, with an annotation specifying "ribbon"; and (6) a scribble outlining a beanie design without a pompom, while annotations and text prompts detailed adding a "Tottenham logo" to the pink-circled area.}
\end{figure}
\clearpage

\end{document}